\documentclass[%
 reprint,
 amsmath,amssymb,
 aps,
pra,
]{revtex4-1}

\usepackage{graphicx}
\usepackage{dcolumn}
\usepackage{bm}
\usepackage{epstopdf}
\usepackage[usenames]{color}
\usepackage{subeqnarray}
\usepackage{subfigure}
\usepackage{easybmat}
\usepackage{amsmath}
\usepackage[colorlinks,
linkcolor=blue, urlcolor=blue, anchorcolor=blue, citecolor=blue]{hyperref}
\usepackage[T1]{fontenc}
\usepackage{mathptmx}
\usepackage{booktabs}

\def\be{\begin{equation}}
\def\ee{\end{equation}}
\def\bea{\begin{eqnarray}}
\def\eea{\end{eqnarray}}
\def\bs{\begin{subequations}}
\def\es{\end{subequations}}

\begin{document}


\title{Distinct Properties of Vortex Bound States Driven by Temperature}

\author{Xinwei Fan}
\author{Xiaoyu Chen}%
\author{Huan Yang}
\author{Hai-Hu Wen}
\email{hhwen@nju.edu.cn}
\address{National Laboratory of Solid State Microstructures and Department of Physics, Center for Superconducting Physics and Materials, Collaborative Innovation Center for Advanced Microstructures, Nanjing University, Nanjing 210093, China}

\begin{abstract}
We investigate the behavior of vortex bound states in the quantum limit by self-consistently solving the Bogoliubov-de Gennes equation. We find that the energies of the vortex bound states deviates from the analytical result $E_\mu=\mu\Delta^2/E_F$ with the half-integer angular momentum $\mu$ in the extreme quantum limit. Specifically, the energy ratio for the first three orders is more close to $1:2:3$ instead of $1:3:5$ at extremely low temperature. The local density of states reveals an Friedel-like behavior associated with that of the pair potential in the extreme quantum limit, which will be smoothed out by thermal effect above a certain temperature even the quantum limit condition, namely $T/T_c<\Delta/E_F$ is still satisfied. Our studies show that the vortex bound states can exhibit very distinct features in different temperature regimes, which provides a comprehensive understanding and should stimulate more experimental efforts for verifications.
\end{abstract}

\maketitle


\section{Introduction}
The vortex predicted by the Ginzburg-Landau theory is a fascinating object which appears in type-II superconductors. Due to the confinement by the superfluid to the quansiparticles within the vortex core, there have been enormous work in exploring the internal structure of this quantized object. The pioneer work of Caroli, de Gennes, and Matricon (CdGM) \cite{RN1416} pointed out that the low-energy bound states in the core of a single vortex in a type II superconductor are discrete and the energy level approximately obeys the relation $\mu\Delta^2/E_F$. However, in most of conventional superconductors, the pair potential $\Delta$ is generally very small compared with the Fermi energy $E_F$, which makes the discrete energy levels hard to be distinguished in experiments and instead usually only one peak consisting of multiple energy levels can be observed at zero bias in the spectrum \cite{RN1432,RN1433,RN1418,RN1434}. The recent scanning tunneling microscopy (STM) measurements carried out on FeTe$_{0.55}$Se$_{0.45}$ \cite{RN1439,RN1419} and FeSe monolayer thin film \cite{RN1437} revealed the discrete energy levels due to a relatively small Fermi energy of these materials \cite{RN1419,RN1435,RN1436}, which enlarges the interval of the energy level and makes it discernable. Apart from that the CdGM states play an important role in thermodynamic and transport properties of the mix state in type II superconductors \cite{RN1388,RN1447,RN1448,RN1449,RN1450}, growing interests have been generated in studying the CdGM states partially due to the possible existence of Majorana zero modes in the vortex cores of topological superconductors\cite{RN1439,RN1442,RN1451,RN1453,RN1454,RN1455}. This may enable the so-called topological superconductor to be an ideal platform for quantum computation\cite{RN1444,RN1443}. 
\par
In addition to the analytical approach \cite{RN1416}, some groups purposed a self-consistent method to solve the Bogoliubov-de Gennes (BdG) equation \cite{RN1411,RN1388}. The pair potential $\Delta(r)$ was found to have an oscillatory behavior with the period $1/k_F$ in the quantum limit  $T/T_c<\Delta/E_F$ \cite{RN1418}. This behavior originates from the oscillatory nature of the quasiparticle wave functions \cite{RN1388} and makes the hypothetical form of the pair potential $\Delta(r)=\tanh{(r/\xi_0)}$ no longer valid in this situation. Thus, it is necessary to carefully re-examine the properties of the vortex bound states in the quantum limit. In this work, we want to address two issues. First, the vortex bound state energies on the quasiparticle spectrum deviate from the linear relation in the quantum limit based on the self-consistent analysis. According to Caroli $et$ $al.$ \cite{RN1416}, the spectrum of the bound states possesses a linear relation, namely $E_\mu=\mu\Delta^2/E_F$, while since the spectrum accumulates near the energy gap, the linear relation is naturally violated for high-lying energy orders. Thus here we only focus on the first three orders, which are also experimentally accessible. Since the angular momentum $\mu$ can only be half integer in the clean limit of a type-II superconductor \cite{RN1418,RN1388,RN1412}, the ratio of the first three orders of the bound state energies should be $1:3:5$. We find that in the extreme quantum limit, the ratio deviates from this ideal value and is in fact temperature dependent. Our calculations show that it is actually quite close to the ratio $1:2:3$ at very low temperatures. Second, our calculation shows a Friedel-like oscillation of the pair potential. Furthermore, the local density of states (LDOS) also exhibits this kind of oscillatory behavior originating from that of the quasiparticle wave functions in the extreme quantum limit. With increasing temperature, this oscillation will be thermally smoothed out.

\section{Bogoliubov-de Gennes equation}
The method of self-consistently solving BdG equation has been well presented in previous works \cite{RN1411,RN1388,RN1412}. However, for consistency we would like still give a brief outline of this process. By utilizing the relation $E_F=\hbar^2k_F^2/2m$ and scaling the length and energy respectively with the coherence length $\xi_0$ and $\Delta_0$ the pair potential far away from the vortex core, we can simplify the parameters characterizing the system by adopting the product of $k_F$ and $\xi_0$ \cite{RN1388}. Thus, the single-particle Hamiltonian takes the dimensionless form $H_0=\frac{-1}{2 k_{F} \xi_{0}} \nabla^{2}-E_{F}$ and we have the BdG equation
\begin{equation}
\label{eq1}
\left[\begin{array}{cc}
H_{0} & \Delta(\mathbf{r}) \\
\Delta^{\star}(\mathbf{r}) & -H_{0}
\end{array}\right]\left(\begin{array}{l}
{u}_n(\mathbf{r}) \\
{v}_n(\mathbf{r})
\end{array}\right)={E}_{n}\left(\begin{array}{ll}
{u}_n(\mathbf{r}) \\
{v}_n(\mathbf{r})
\end{array}\right),
\end{equation}
where $u_j(\mathbf{r})$ and $v_j(\mathbf{r})$ are the quasiparticle wave functions, $u^2$ and $v^2$ reflect the un-occupation and occupation probability of the Cooper pairs, all these should be self-consistently obtained with the condition

\begin{equation}
\label{eq2}
\Delta(\mathbf{r})=g \sum_{\left|E_{n}\right| \leq \omega_{D}} u_{n}(\mathbf{r}) v_{n}^{*}(\mathbf{r})\left\{1-2 f\left(E_{n}\right)\right\},
\end{equation}
where g is the coupling strength, $f(E)$ the Fermi function, and $\omega_D$ the Debye frequency functioning as the energy cutoff. With properly choosing the gauge of $\Delta(\mathbf{r})$, its phase can be cancelled out by that of the quasiparticle wave function \cite{RN1388,RN1412}. Since the system possesses a cylindrical symmetry, we can expand the wave functions with the Bessel function $J_m(r)$ as
\begin{equation}
\begin{array}{l}
u_{n}(r)=\sum_{j} c_{n j} \phi_{j \mu-1 / 2}(r) \\
v_{n}(r)=\sum_{j} d_{n j} \phi_{j \mu+1 / 2}(r),
\end{array}
\end{equation}
where $\phi_{j m}(r)=\frac{\sqrt{2}}{R J_{m+1}\left(\alpha_{j m}\right)} J_{m}\left(\alpha_{j m} r/R\right)$ with $\alpha_{jm}$ the $i$th zero point of $J_m(r)$ and $j=1, \ldots, N$. Now we should give an initial $\Delta(r)$, put it into Eq. \ref{eq1}, solve the $2N\times2N$ eigenvalue problem, reproduce the quasiparticle wave functions and put them into Eq. \ref{eq2} to generate a new $\Delta(r)$. We obtain the true wave functions once $\Delta(r)$ converges.

\begin{figure}
    \centering
    \includegraphics[width=6.3cm]{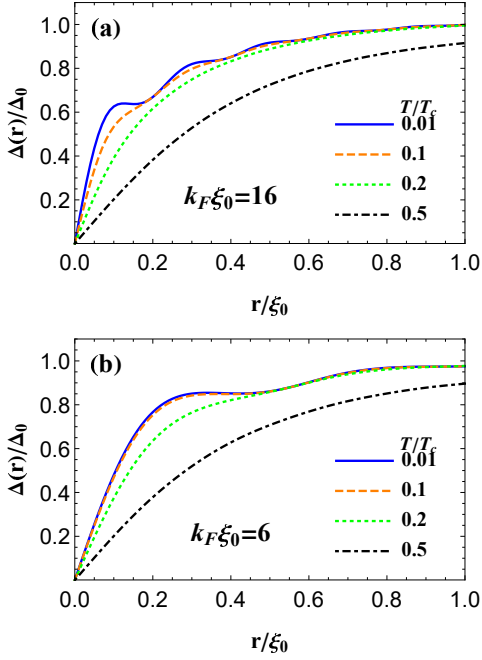}
    \caption{Color online. Spatial dependence of the pair potential $\Delta(r)$ for different temperatures. The characteristic parameters $k_F\xi_0$ for (a) and (b) are $16$ and $6$, respectively.}
    \label{fig1}
\end{figure}

\section{Analysis}
After obtaining the quasiparticle wave functions, we can inspect the physical quantities that we are interested in. Here we focus on the low-energy excitations and the local density of states $\sum_{n}\left[\left|u_{n}(\mathbf{r})\right|^{2} f^{\prime}\left(E-E_{n}\right)+\left|v_{n}(\mathbf{r})\right|^{2} f^{\prime}\left(E+E_{n}\right)\right]$ of the bound states, which can be directly observed with STM experiments. Figure \ref{fig1} shows the obtained pair potential $\Delta(r)$ with two $k_F \xi_0$ by self-consistent solutions to the BdG equations at different temperatures. We can see that $\Delta(r)$ exhibits an oscillatory behavior at very low temperatures, this has actually been reported in previous works \cite{RN1411,RN1388}. Since the period of the oscillation is proportional to $1/k_F$ \cite{RN1388}, the oscillation is more significant with a larger $k_F$.
\par
\begin{figure}
    \centering
    \includegraphics[width=7.3cm]{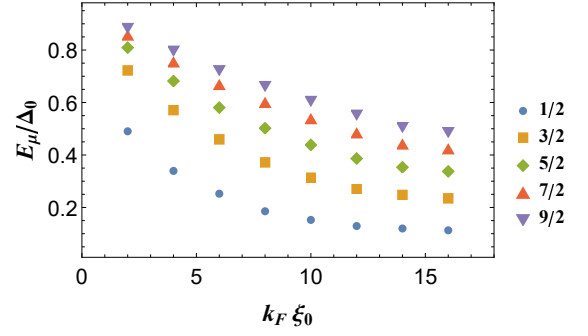}
    \caption{Color online. The first five bound states energy $E_\mu$ at different characteristic parameter $k_F\xi_0$.}
    \label{fig2}
\end{figure}
As for the low-energy excitations, the spectra are shown in Fig. \ref{fig2}.
\begin{figure}
    \centering
    \includegraphics[width=6.3cm]{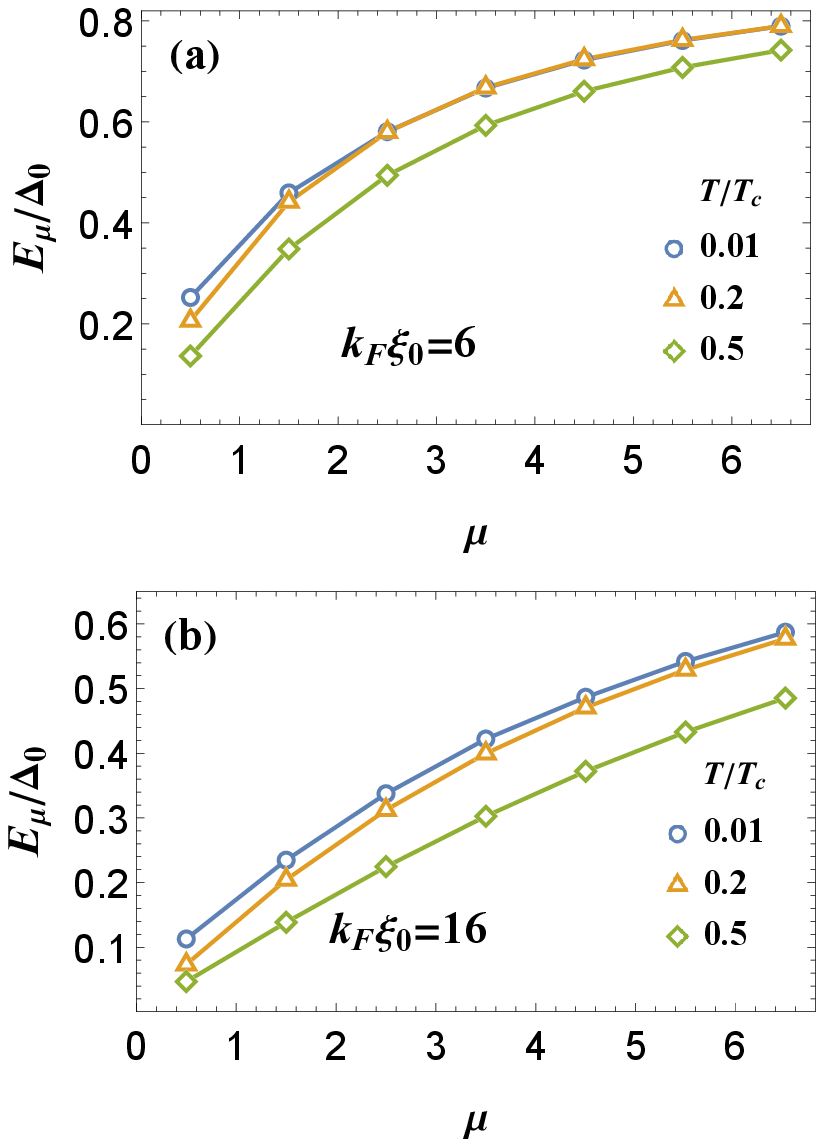}
    \caption{Color online. The quasiparticle spectrum in different temperatures. The $k_F\xi_0$ adopted in (a) and (b) are 16 and 6, respectively. The deviation from the ideal linear relation becomes more evident with a smaller $k_F\xi_0$ as shown in (a).}
    \label{fig3}
\end{figure}
According to Caroli $et$ $al.$ \cite{RN1416}, the discrete energy level possesses a linear relation, that is $E_\mu = \mu \Delta^2/E_F $. Allowing for that the angular momentum $\mu$ can only take the values of half integers, the energy ratio for the lowest three excitations should be $1:3:5$. However, by solving the BdG equation self-consistently, we find that the bound states behave differently with varying temperatures even below the quantum limit. At extremely low temperature, the ratio significantly deviates from the ideal value $1:3:5$. Figure \ref{fig3} shows the excitation energies versus the angular momentum $\mu$. We can see that with a larger $k_F\xi_0$, the dispersion is closer to a linear relation, while for a smaller one, the deviation from the linear relation becomes distinct since the minimum excitation energy is larger in this situation leading to a faster convergence towards the energy gap as shown in Fig. \ref{fig2}. 
\begin{table}[htbp]
	\centering
	\setlength{\tabcolsep}{8mm}{
	\caption{The energy ratio of the first three orders of the spectrum with different characteristic parameters $k_F\xi_0$ and temperatures.}
	\label{table1}
	\begin{tabular}{ccc}
		\toprule
		$k_F \xi_0$&$T/T_c$&energy ratio \\
		\midrule
		&0.01&$1:1.82:2.31$ \\
		6&0.2&$1:2.14:2.81$\\
		&0.5&$1:2.55:3.62$\\
		\hline
		\\
		&0.01&$1:2.08:2.99$\\
		16&0.2&$1:2.78:4.25$\\
		&0.5&$1:2.96:4.79$\\
		\hline
		\\
		&0.01&$1:2.12:3.12$\\
		20&0.2&$1:2.87:4.51$\\
		&0.5&$1:2.98:4.89$\\
		\bottomrule
	\end{tabular}}
\end{table}
Table \ref{table1} gives the energy ratio based on our calculations with different $k_F\xi_0$ and temperatures. The ratio of the vortex bound state energies clearly deviate from $1:3:5$ at low temperatures especially for a small $k_F\xi_0$. While for a relatively large $k_F\xi_0$, at extremely low temperature $T=0.01T_c$, the ratio is very close to $1:2:3$ and it approaches the ideal value $1:3:5$ at relatively high temperature $T=0.5T_c$.
\begin{figure}
    \centering
    \includegraphics[width=8.2cm]{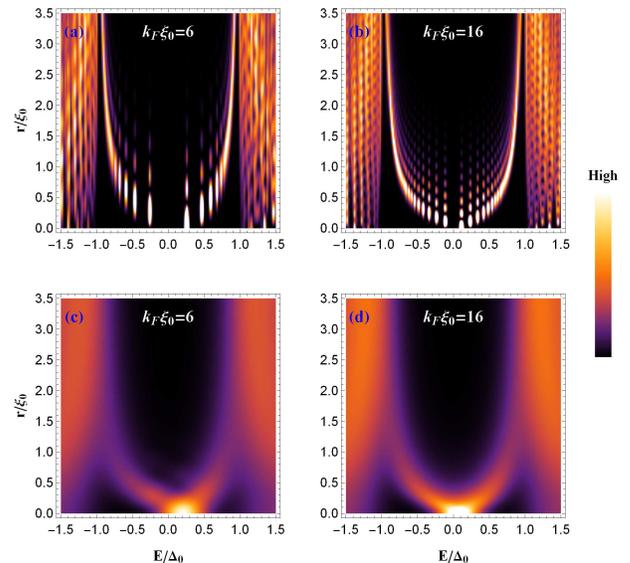}
    \caption{Color online. (a) and (b) are the spatial evolution of the local density of states at $T=0.01T_c$. (c) and (d) are the corresponding LDOS at $T=0.2T_c$. The characteristic parameter $k_F\xi_0$ is 6 in (a) and (c), and 16 in (b) and (d). }
    \label{fig4}
\end{figure}
\par
\begin{figure}
    \centering
    \includegraphics[width=9.1cm]{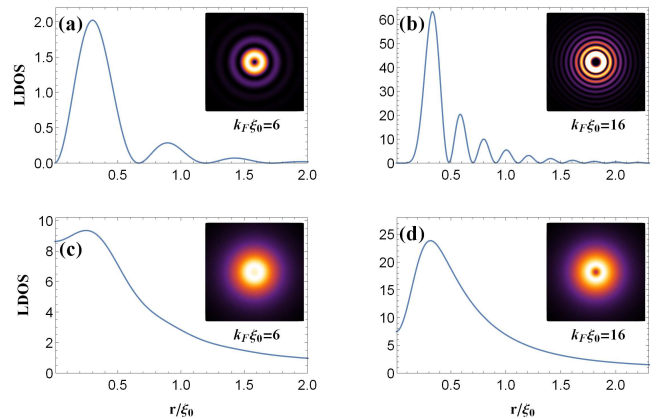}
    \caption{Color online. The energy cut $E=0.5\Delta_0$ of Fig. \ref{fig4}. (a) and (b) show clearly the Friedel-like oscillation of the LDOS. The oscillation is smoothed out due to the thermal smearing as shown in (c) and (d). The insets show the corresponding two-dimensional LDOS.}
    \label{fig5}
\end{figure}
Since STM is currently the most suitable way to experimentally study the bound states in the vortex core and there have already been some works focusing on this subject \cite{RN1419,RN1420,RN1421,RN1422,RN1423}, we gives the LDOS for two characteristic parameters $k_F\xi_0=6$ and $k_F\xi_0=16$ in Fig. \ref{fig4}. At extremely low temperature $T=0.01T_c$, it is very easy to discern the LDOS peaks for both $k_F\xi_0$ as shown in Figs. \ref{fig4}(a) and \ref{fig4}(b). At a moderate temperature $T=0.2T_c$, the LDOS is heavily smeared, which makes the LDOS peaks no longer distinguishable as shown in Figs \ref{fig4}(c) and \ref{fig4}(d). We can also observe a spatial oscillation in Figs. \ref{fig4}(a) and \ref{fig4}(b). Figure \ref{fig5} gives the spatial dependence of the LDOS at the energy $E=0.5\Delta_0$ with the same $k_F\xi_0$ and temperatures adopted in Fig. \ref{fig4}. The oscillatory behavior of the LDOS is presented in Figs. \ref{fig5}(a) and \ref{fig5}(b). At $T=0.2T_c$, the oscillation disappears due to the thermal smearing. Thus, in order to observe the Freidel-like oscillations of the LDOS in experiments, it is essential to do the experiments at extremely low temperatures when the extreme quantum limit condition is satisfied. Our results clearly illustrate that the vortex bound state energies and the related spectrum are strongly dependent on the temperature. The ratio between the bound state energies can deviate from the analytical expectation $1:3:5$. At extreme low temperatures, both the pairing potential and LDOS exhibit the Freidel-like oscillations. Evidences for these predictions can be found in recent experiments carried out on iron-based superconductors \cite{RN1439,RN1419,RN1437,chen2021friedel}. Especially, both the deviation from the energy ratio $1:3:5$ and the Friedel-like oscillation of the LDOS have been observed in recent STM experiment in KCa$_2$Fe$_4$As$_4$F$_2$ \cite{chen2021friedel}. Nevertheless, more experimental verifications in other superconductors with a relatively large ratio of $\Delta /E_F$ are still desired.

\section{Summary}
In conclusion, by self-consistently solving the BdG equations, we find that the vortex bound state energies of the spectrum deviate from the analytically expected result $E=\mu\Delta^2/E_F$ in the quantum limit due to the oscillatory behavior of the pair potential. Especially, the energy ratio of the first three orders of the spectrum significantly deviates from $1:3:5$, but is close to $1:2:3$ at extremely low temperatures. Similarly, the local density of states also exhibit this kind of spatial oscillatory behavior in the quantum limit, which can be directly observed with STM experiments. With increasing temperature, this oscillation are smeared out by thermal effect. Our work provides a comprehensive understanding of the vortex bound states based on self-consistent solutions to the BdG equations.
\section*{Acknowledgments}
We thank Da Wang and Christopher Berthod for very helpful discussions. This work was supported by National Key R\&D Program of China (Grants No. 2016YFA0300401), National Natural Science Foundation of China (No. 12061131001), and the Strategic Priority Research Program (B) of Chinese Academy of Sciences (Grants No. XDB25000000).
%

\end{document}